\documentclass[aps,reprint,amsmath,amssymb,eqsecnumaps,pra]{revtex4-2}
\usepackage{graphicx}
\usepackage{dcolumn}
\usepackage{bm}
\usepackage{float}
\usepackage{xcolor}
\usepackage{lineno}
\usepackage{verbatim} 
\usepackage[normalem]{ulem}
\usepackage{lineno}

\newcommand{\p}{\partial}

\newcommand{\ep}{\varepsilon}
\newcommand{\vta}{\vartheta}
\newcommand{\om}{\omega}

\newcommand{\vD}{\varDelta}
\newcommand{\nn}{\nonumber}
\newcommand{\ta}{\theta}

\newcommand{\wh}{\widehat}

\newcommand{\cH}{{\cal H}}
\newcommand{\cF}{{\cal F}}

\newcommand{\cW}{{\cal W}}

\newcommand{\be}{\begin{equation}}                                       
	\newcommand{\ee}{\end{equation}}
\newcommand{\ba}{\begin{eqnarray}}
	\newcommand{\ea}{\end{eqnarray}}
\newcommand{\bref}[1]{(\ref{#1})}
\newcommand{\bsub}{\begin{subequations}}                      
	\newcommand{\esub}{\end{subequations}}     

\newcommand{\bi}[1]{\bibitem{#1}}\newcommand{\lab}[1]{\label{#1}}

\begin{document}
	
	\preprint{APS/123-QED}
	
	\title{
Ladder of Eckhaus instabilities and parametric conversion in 
chi(2)~microresonators
	}
	\author{Danila N. Puzyrev}
	\author{Dmitry V. Skryabin}
	\email{d.v.skryabin@bath.ac.uk}
	\affiliation{Department of Physics, University of Bath, Bath BA2 7AY, UK}%
	
	\begin{abstract}
{\bf Abstract.} 
	Low loss microresonators have revolutionised nonlinear and quantum optics over the past decade. In particular,  microresonators with the second order, chi(2), nonlinearity have the advantages of broad spectral tunability and low power frequency conversion. Recent observations have highlighted that the parametric frequency conversion in chi(2) microresonators is accompanied by stepwise changes in the signal and idler frequencies. Therefore, a better understanding of the mechanisms 
and development of the theory underpinning this behaviour is  timely. Here, we report that the stepwise frequency conversion originates from the discrete sequence of the so-called Eckhaus instabilities. After discovering these instabilities in fluid dynamics in the 1960s, they have become a broadly spread interdisciplinary concept. Now, we demonstrate that the Eckhaus mechanism also underpins the ladder-like structure of the frequency tuning curves in chi(2) microresonators.
	\end{abstract}
	\maketitle

	\maketitle

\noindent{\bf Introduction.} Nonlinear and quantum optics in the ring microresonators have attracted a great deal of attention over the past few years~\cite{di1,st1}. 
Its applications include classical and quantum information processing~\cite{rf1,w1,w4}, precession spectroscopy~\cite{di1} and exoplanet discovery~\cite{pl}.
While the third-order nonlinear response (Kerr effect) of microresonators remains the workhorse of research in this area, the second-order, $\chi^{(2)}$, nonlinearity is attracting an increased attention~\cite{st1,prl,tang1}.
$\chi^{(2)}$ response is generally stronger than Kerr and offers greater flexibility with the spectral tunability and access to a range of quantum effects~\cite{w4,fabr,ang,gol}. The respective material and fabrication issues have matured significantly~\cite{optr}, and  the phase-matching by the  
self-induced gratings in Si$_3$N$_4$ have been demonstrated~\cite{kart,bres,papp}.

With all the innovations in mind, the temperature, $T$, control of the refractive index to provide the signal-idler-pump energy conservation, $\hbar\om_{s}+\hbar\om_{i}=\hbar\om_p$, remains an essential method of the frequency tuning of the down-converted signal in microresonators since their very early days~\cite{byer0,byer1} and till now~\cite{opo1,opo3}. For the total (material+geometry) resonator dispersion being well approximated by the second-order expansion, the index matching curves, $T$ vs $\om_{s,i}$, take the parabolic shape. It was demonstrated already in Ref.~\cite{byer1}, and then has become a staple for many results on frequency conversion in $\chi^{(2)}$ microresonators, see, e.g.,~\cite{opo3,opo2,st0} and references therein. 
Recent parametric down-conversion results in the 
thin-film quasi-phase-matched~\cite{opo2} and whispering gallery~\cite{opo4} LiNbO$_3$ microresonators have also used the pump laser frequency as a control parameter 
leading to the parabolic $\om_p$ vs $\om_{s,i}$  tuning curves~\cite{opo2,opo4}. In what follows, we develop
a theory underpinning the pump-frequency-tuning approach.

A  distinct feature of the tuning process in microresonators is that the signal and idler frequencies vary discretely, i.e., with a jump, when one signal-idler pair switches to the next in sequence~\cite{opo3,opo2,opo4}. This discreteness raises some interesting questions. E.g., is there a bifurcation scenario leading to the stepwise mode-number switch, or this is a bifurcation free parameter-drag effect, and what could be the locking interval in the parameter space providing the device operation in the desired pair of the resonator modes? Our theory answers the above questions and reveals that the instability scenario, or rather a ladder-like sequence of the instabilities, behind the tuning of the parametric frequency conversion in microresonators is known in the contexts of fluid dynamics and pattern formation as the Eckhaus instability~\cite{rmp,rmp0,ec4,gom}.
The term of the Eckhaus, or so-called secondary, instability is used to describe the bifurcation that changes  the period of a nonlinear wave~\cite{rmp,rmp0,ec4,gom,longhi2,taki,ec2,pere}.
We also find an explicit condition on the power vs loss balance which explains the cutting of the optical parametric oscillator (OPO) tuning parabola at one or both of its ends and limits the achievable spectral separation between the signal and idler photons.

\vspace{1mm}
\noindent{\bf Results and discussion}

\vspace{1mm}
\noindent{\bf Model and classification of OPO regimes.} 
Frequencies of the modes in high-finesse resonators are rigidly linked to the quantised set of the wavenumbers. Therefore, the step in frequency also implies  the  abrupt change of the spatial period, as it happens in the Eckhaus instability scenario. Accounting for the quantisation of the spectrum calls for the model formulation and methodology different from the continual models of transverse nonlinear optics~\cite{longhi2,oppo,st,opomy,tl0,tl1}, fibre-loop~\cite{taki,ec2,mar} and bow-tie~\cite{mar2,leo1,bab} cavities, cf. with the soliton and frequency conversion theories in the high-finesse microresonators~\cite{che,kip,prap,men,josab,pra3,pra2,prr,ol3}.
\begin{figure}[t]
	\centering{	
		\includegraphics[width=0.48\textwidth]{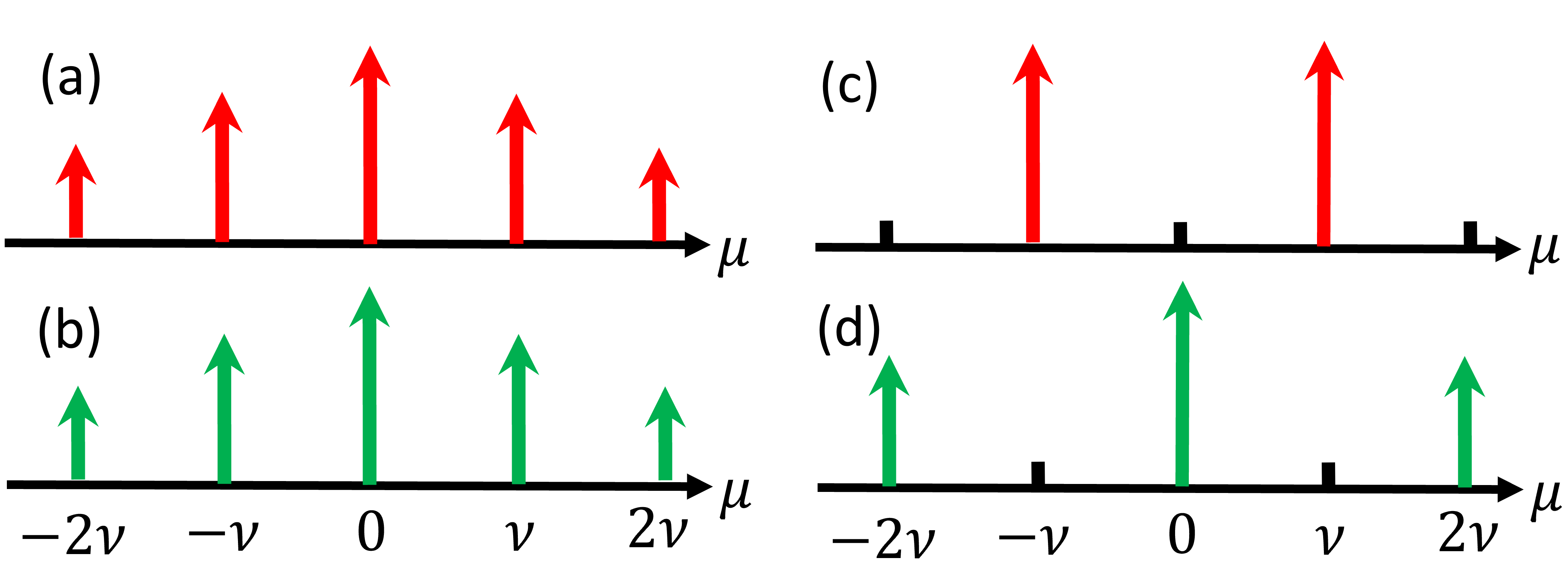}	
	}
	\caption{{\bf Generic and staggered frequency combs.} {\bf a,b} is an illustration of the generic comb with the spatial period $2\pi/\nu$. {\bf c,d} is the staggered comb with the period $\pi/\nu$. Red and green colours mark the signal and pump combs, respectively. The resonator parameters~\cite{prl}  are:
		linewidths: $\kappa_a/2\pi=1$MHz, $\kappa_b/2\pi=2$MHz; 
		repetition rates: $D_{1a}/2\pi=21$GHz, $D_{1b}/2\pi=20$GHz;
		dispersions: $D_{2a}/2\pi=-100$kHz, $D_{2b}/2\pi=-200$kHz;
		nonlinear coefficients: $\gamma_{a,b}/2\pi=300\text{MHz}/\sqrt{\text{W}}$.}
	\lab{f0}
\end{figure}

We now set the model in the transparent and rigorous manner.
We assume that the  pump laser frequency, $\om_p$, is tuned around the frequency, $\om_{0b}$, of the resonator mode with the number $2M$. $2M$ equals the number of wavelengths fitting along the  ring circumference. We also express the multimode intra-resonator electric field
centred around $\om_p$ and its half-harmonic, $\om_p/2$, as 
\be
\begin{split}
&e^{iM\vta-i\frac{1}{2}\om_p t}\sum_\mu a_\mu e^{i\mu\ta}+c.c.,\\ 
&e^{i2M\vta-i\om_p t}\sum_\mu b_\mu e^{i\mu\ta}+c.c.,~~\ta=\vta- D_{1}t,
\end{split}
\lab{field}
\ee
respectively. Here, $\vta=(0,2\pi]$ is the angular coordinate, and
$\ta$ is its transformation to the reference frame rotating with the  rate $D_{1}/2\pi$. 
'$a$' and '$b$' mark the half-harmonic, i.e., signal, and pump fields, respectively.
$\mu=0,\pm 1,\pm 2,\dots$ is the relative mode number, and the resonator frequencies are
\be
\om_{\mu \zeta}=\om_{0\zeta}+\mu D_{1\zeta}+\tfrac{1}{2}\mu^2 D_{2\zeta},~\zeta=a,b,
\lab{om}\ee
where,  $D_{1\zeta}/2\pi$ are the free spectral ranges, FSRs, and $D_{2\zeta}$ are dispersions. In what follows, we choose $D_1=D_{1a}$. 
The frequency, i.e., phase, matching parameter for the non-degenerate parametric process is defined as
\ba
\ep_\mu&=&\om_{\mu a}+\om_{-\mu a}-\om_{0b}
\nn\\
&=&\frac{c}{R}\left[
\frac{M+\mu}{n_{M+\mu}}+
\frac{M-\mu}{n_{M-\mu}}
-\frac{2M}{n_{2M}}
\right].
\label{ep}
\ea 
Here, $n_m$  is the effective refractive index taken for the frequencies of the  modes with the absolute numbers $m=M\pm\mu$ (signal and idler) and $2M$ (pump), $c$ is the vacuum speed of light and $R$ is the resonator radius. 
E.g., $\ep_0=2\om_{0a}-\om_{0b}=0$ corresponds to the exact matching for the degenerate parametric conversion, $n_M=n_{2M}$.

Coupled-mode equations governing the evolution of $a_\mu(t)$, $b_\mu(t)$ are~\cite{tang1,prap,josab}
\be
\lab{tp1}
\begin{split}
	i\p_t a_{\mu}=\delta_{\mu a}a_{\mu} -& \frac{i\kappa_a}{2}
	a_{\mu}
	\\ -&\gamma_a\sum_{\mu_1 \mu_2}\wh\delta_{\mu,\mu_1-\mu_2}b_{\mu_1}a^*_{\mu_2},
		\\
	i\p_tb_{\mu}=\delta_{\mu b}b_{\mu} - &\frac{i\kappa_b}{2}
	\big(b_{\mu}-\wh\delta_{\mu,0}\cH\big)
	\\	-&\gamma_b\sum_{\mu_1 \mu_2}\wh\delta_{\mu,\mu_1+\mu_2}a_{\mu_1}a_{\mu_2},
\end{split}
\ee
where $\wh\delta_{\mu,\mu'}=1$  for $\mu=\mu'$ and is zero otherwise.
$\cH$ is the pump parameter,
$\cH^2=\cF\cW/2\pi$, where
$\cW$ is the laser power, and $\cF=D_{1b}/\kappa_{b}$ is finesse~\cite{josab}.  
$\delta_{\mu\zeta}$ are the detuning parameters in the rotating reference frame,
$\delta_{\mu a}  =(\om_{\mu a}-\tfrac{1}{2}\om_p)-\mu D_{1a}$ and $
	\delta_{\mu b}  =(\om_{\mu b}-\om_p)-\mu D_{1a}$.
$\delta_{0b}$ is the pump detuning which is 
the main control parameter~\cite{opo3,opo2,opo4}.
The parameter values used to scale our results to physical units are listed in the caption of Fig.~\ref{f0}.

Simple algebra reveals  that all modal detunings in the half-harmonic
signal can be expressed via $\delta_{0b}$ and the respective phase matching parameters,
\be
\delta_{\mu a}=\om_{0a}-\tfrac{1}{2}\om_p+\tfrac{1}{2}\mu^2D_{2a}=\tfrac{1}{2}\delta_{0b}+
\tfrac{1}{2}\ep_\mu.
\lab{fi1a}
\ee
While $\delta_{\mu a}$ and $\ep_\mu$ do not depend on the repetition rate difference, $\delta_{\mu b}$ does,
\be
	\delta_{\mu b}  =
	\delta_{0 b}+\mu( D_{1b}-D_{1a})+\tfrac{1}{2}\mu^2 D_{2b}.
\lab{fi1b}
\ee

Depending on the  pump power, the classical half-harmonic signal can
be either zero or not. This is reflected in the structure of Eq.~\bref{tp1} and its solutions. 
Three types of solutions we should highlight are
\ba
	&&\text{(i)~no-OPO~state:~}
	\nn\\
	&& a_{\mu}=0, b_0=\frac{-i\kappa_b\cH}{2\delta_{0b}-i\kappa_b}, b_{\mu\ne 0}=0; 
\lab{p1}\\
	&&\text{(ii)~degenerate~OPO~state:~}
	\nn\\
	&&a_0\ne 0, b_0\ne 0, a_{\mu\ne 0}=0, b_{\mu\ne 0}=0; 
\lab{p2}
\ea
and a family of 
\ba
	&&\text{(iii)~non-degenerate~OPO~states:~}
	\nn\\
	&&a_{\pm\nu}\ne 0, b_0\ne 0, a_{|\mu|\ne |\nu|}\approx 0, b_{\mu\ne 0}\approx 0.
\lab{p29}
\ea
Though Eq.~\bref{tp1} do not have a closed analytical solution for the non-degenerate OPO, the experimental data demonstrate that the states with the  $|b_0|^2$ and $|a_{\pm\nu}|^2$ powers strongly dominating across the whole spectrum both exists and can be tuned to change from one $\nu$ to the other, and therefore, they
represent the practically desirable regimes of the microresonator operation~\cite{opo1,opo3,opo2,st0}. 
The explicit expressions for the non-zero modes in Eqs.~\bref{p2}, \bref{p29} are introduced later.

In  the fluid dynamics context~\cite{rmp,rmp0,ec4}, the transition from the no-OPO to an OPO state would correspond to the Benjamin-Feir instability (emergence of the signal), while moving from the OPO state operating in the $\pm\nu$ pair of modes to the $\pm(\nu+1)$ pair would be the Eckhaus instability, see, e.g., Ref.~\cite{rmp} that elucidates the difference between the two. 
The general form of Eq.~\bref{tp1} is not tractable analytically regarding the issue of the interplay between the different OPO states, and, therefore, in the next section, we derive the reduced model that allows such analysis to be carried out in a transparent form.
 
Apart from the OPO regimes listed above, Eq.~\bref{tp1} allow for 
the multi-mode frequency comb solutions, that could be either stationary or time-dependent. The left column of Fig.~\ref{f0} schematically illustrates  a solution of Eq.~\bref{tp1} corresponding to the {\em generic} frequency comb with the spatial period $2\pi/\nu$. The structure of Eq.~\bref{tp1} also admits a family of the spectrally {\em staggered} combs,  see the right column in Fig.~\ref{f0}, and we will show below that the non-degenerate OPO states are, in fact, approximations of the staggered combs.

\begin{figure*}[t]
	\centering{	
		\includegraphics[width=\textwidth]{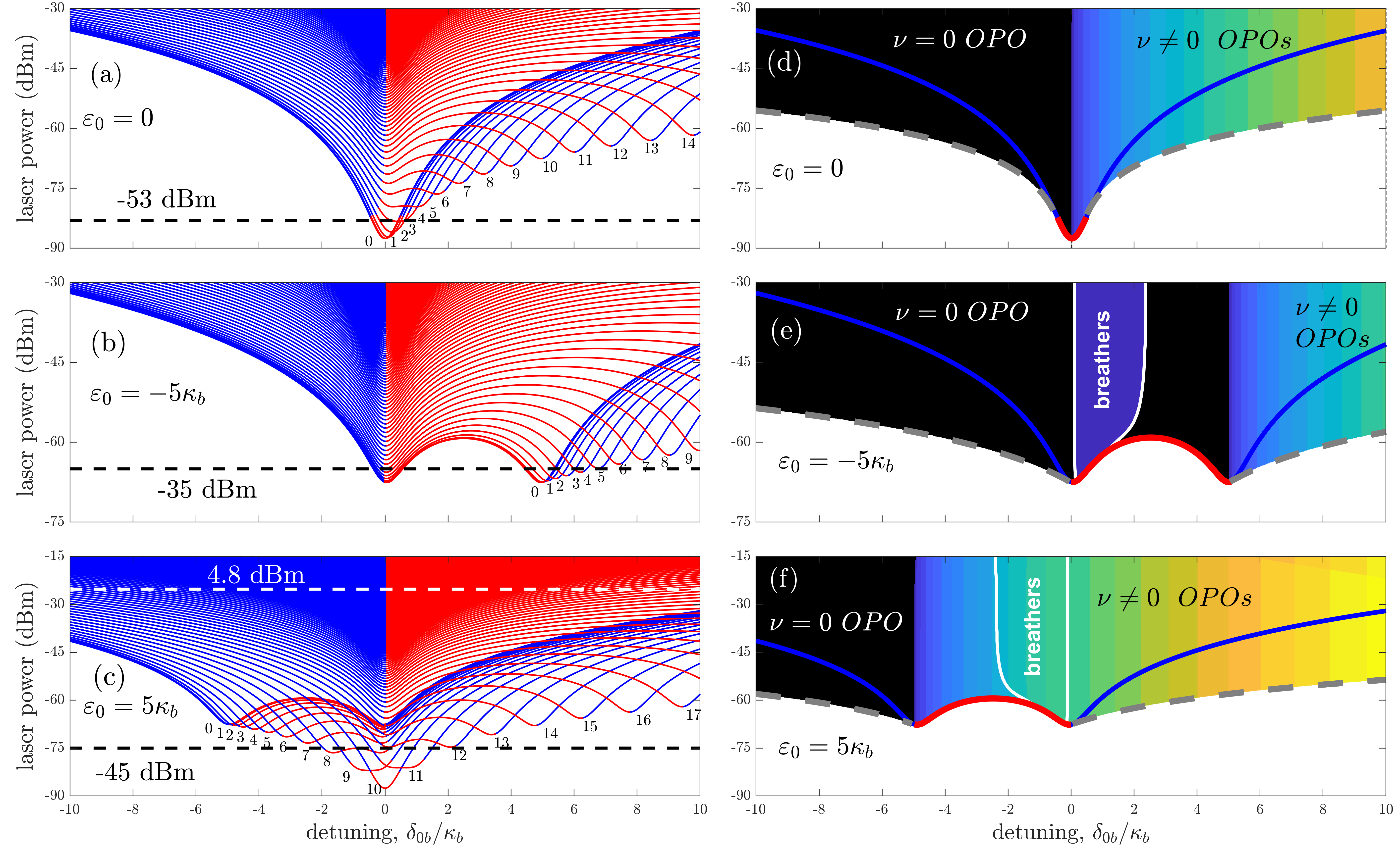}	
	}
	\caption{{\bf Stability ranges of the degenerate optical parametric oscillator (OPO) and no-OPO   states}. {\bf a,b,c} Instability boundaries  of the no-OPO state relative to the excitation of the sideband pairs $\pm\mu$ (Benjamin-Feir instability), where $\mu$'s are indicated.   Red and blue lines  correspond to the super- and sub-critical bifurcations, respectively. {\bf d,e,f} The existence and stability ranges of the degenerate, $\nu=0$, OPO state. The black colour marks the range of the stable operation of the degenerate OPO.  The discretely varying blue-to-orange colour-map shows the range of the non-degenerate OPOs. The full white lines delimit the range of the oscillatory (Hopf) instabilities leading to the breathing combs. 
		Red and blue lines  correspond to the super- and sub-critical bifurcations, respectively, see Eq.~\bref{q0}. The grey dashed line marks Eq.~\bref{lim}.}
	\lab{f2}
\end{figure*}

\begin{figure*}[t]
	\centering{	
		\includegraphics[width=\textwidth]{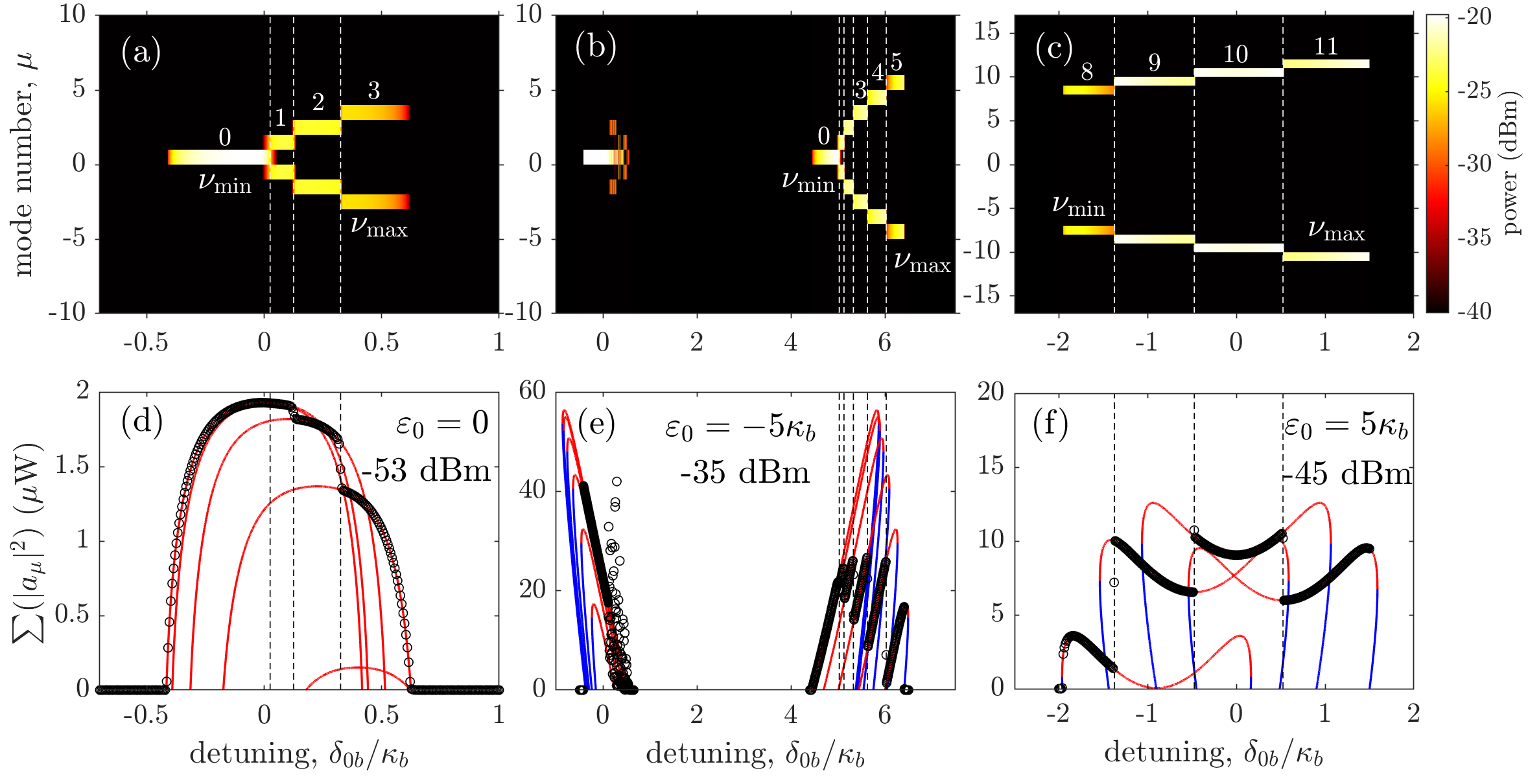}	
	}
	\caption{ {\bf  Eckhaus instability ladders for small laser powers.} Signal mode numbers (a,b,c), i.e., the ladder of Eckhaus instabilities,  and  power of the generated sidebands (d,e,f) vs detuning for the pump powers shown by the black-dashed lines in the left column of Fig.~\ref{f2}: (a,d)~$\cW=5$nW ($-53$dBm), $\ep_0=0$; 
		(b,e)~$\cW=316$nW ($-35$dBm), $\ep_0=-5\kappa_b$; (c,f)~$\cW=31.6$nW ($-45$dBm), $\ep_0=5\kappa_b$.
		{\bf a,b,c}  Numerically simulated, Eq.~\bref{tp1}, excitation of the signal sidebands (field '$a$') vs pump detuning, $\delta_{0b}$. The numbers inside the panels indicate the respective sideband orders. The vertical dashed lines mark the instability boundaries of the states $\nu$ relative to the excitation of the $\nu+1$ sideband pair, i.e., 
		$\lambda_{\nu,\nu+ 1}=0$ conditions, see Eqs.~\bref{th1}, \bref{in1}. 
		The colour scale shows the sideband power.
		{\bf d,e,f}
		Red and blue lines show the sideband powers vs $\delta_{0b}$ as given by the analytical solutions in Eq.~\bref{x3a}:  $|a_{\nu}^+|^2+|a_{-\nu}^+|^2$ (red) and $|a_{\nu}^-|^2+|a_{-\nu}^-|^2$ (blue). 
		Black circles show the numerically computed sideband powers corresponding to the data in the top row. The detuning range around $\delta_{0b}\approx 0.5\kappa_b$ in the 
		middle column corresponds to the breather states.
	}
	\lab{f3}
\end{figure*}

\begin{figure*}[t]
	\centering{	
		\includegraphics[width=\textwidth]{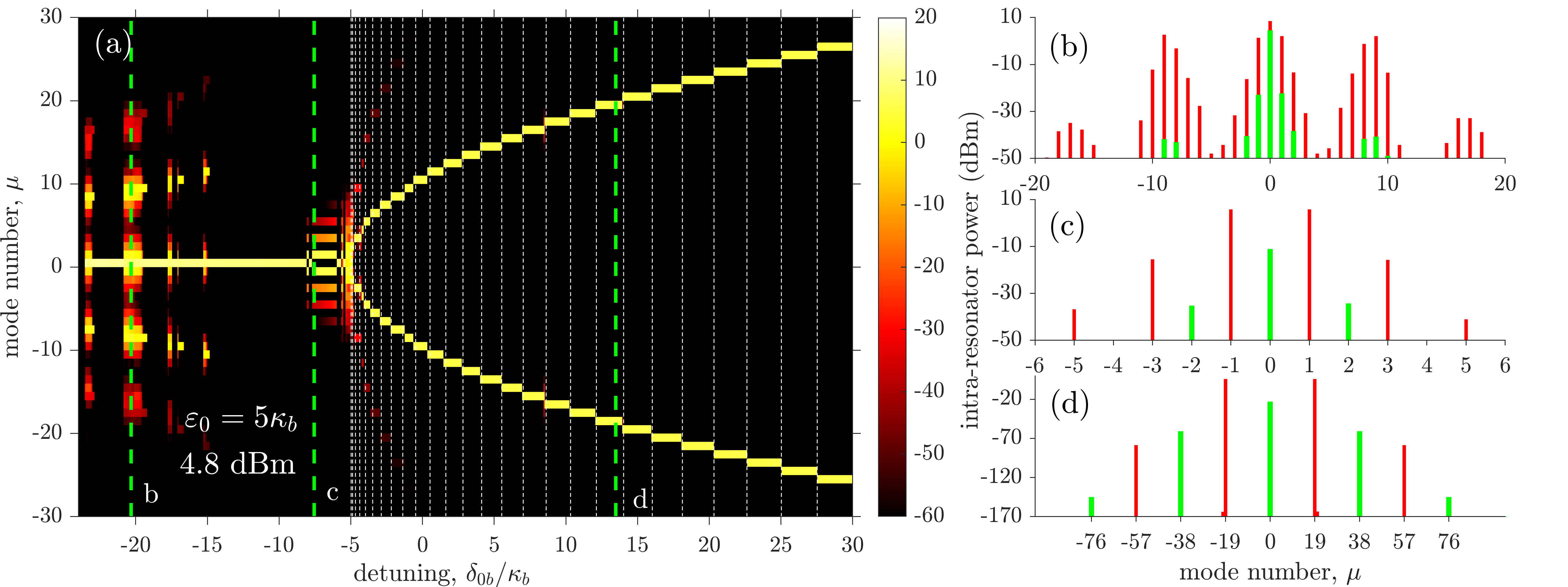}	
	}
	\caption{{\bf Eckhaus instability ladder for high laser power.} {\bf a} Ladder of Eckhaus instabilities for high powers.  Numerically computed, Eq.~\bref{tp1}, spectra of the optical parametric oscillator (OPO) signal 
		vs pump detuning, $\delta_{0b}$, for the pump laser power $\cW=3$mW  ($4.8$dBm) and  $\ep_0=5\kappa_b$, see the horizontal white dashed line in Fig.~\ref{f2}a. The vertical dashed lines map the sequence of the Eckhaus instabilities, $\lambda_{\nu,\nu\pm 1}=0$, see Eqs.~\bref{th1}, \bref{in1}. 
		The colour scale shows the sideband power in dBm. {\bf b,c,d}  show the details of the spectra of the pump (green) and OPO signal (red) corresponding to the dashed green lines in (a). Spectra in (c) and (d) are the staggered combs. (d) corresponds to the non-degenerate OPO state that is very well approximated by Eq.~\bref{x3a}. The vanishing power levels have been included in (d) in order to demonstrate, that the non-degenerate OPO states belong to the staggered comb family.  }
	\lab{f4}
\end{figure*}

\begin{figure}[t]
	\centering{	
		\includegraphics[width=0.48\textwidth]{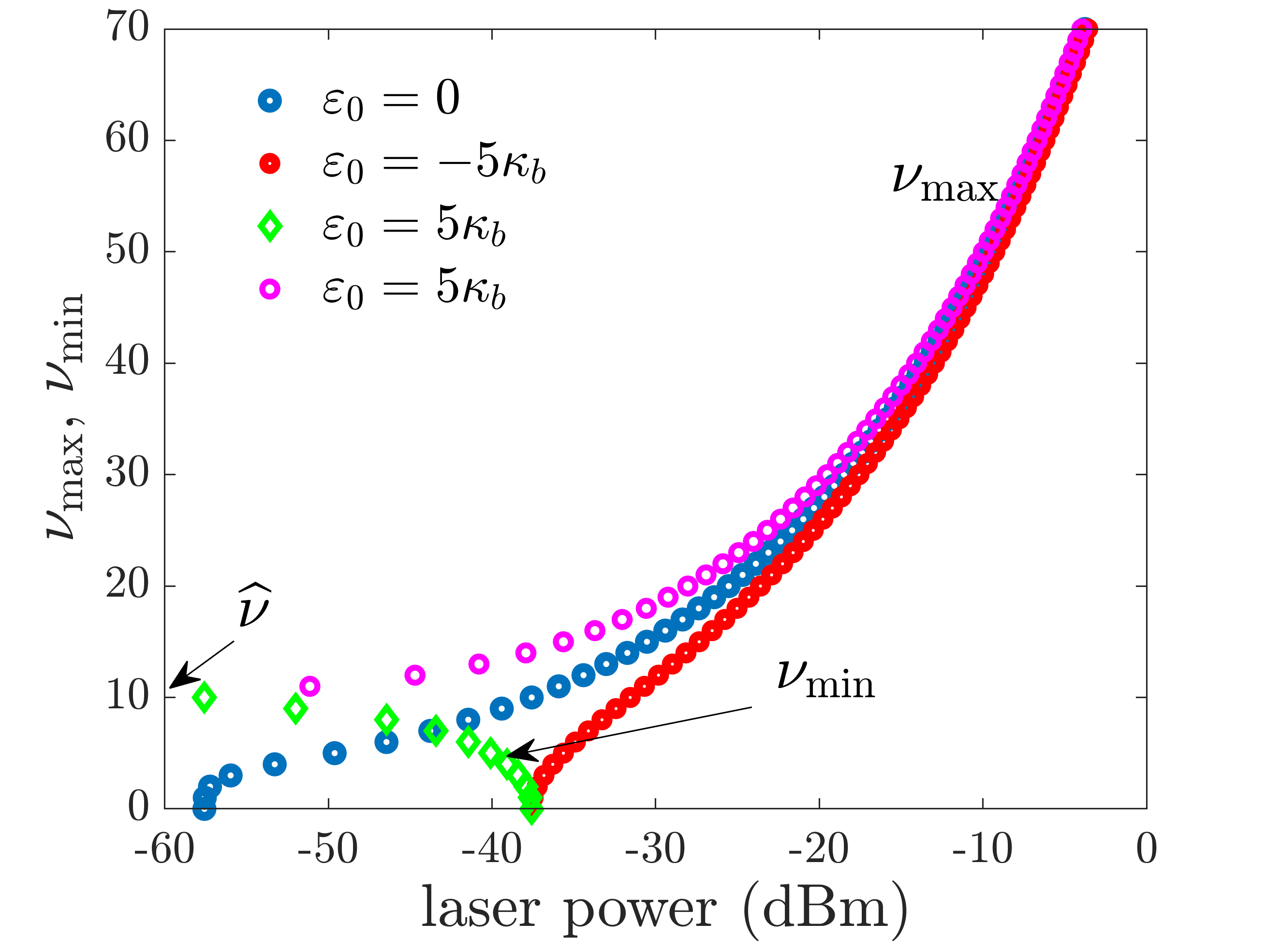}	
	}
	\caption{{\bf Maximal and minimal optical parametric oscillator (OPO) sideband orders achievable by the pump frequency scan.} The pump power dependencies of the maximal, $\nu_{\max}$, and minimal, $\nu_{\min}$, mode numbers corresponding to the $\pm\nu$ OPO states that can be generated by the pump frequency tuning method. $\nu_{\min}$ is different from zero only for $\ep_0>0$. $\wh\nu\approx\sqrt{-\ep_0/D_{2a}}$ corresponds to the lowest possible excitation threshold for $\ep_0>0$, see text between Eqs.~\bref{q0} and \bref{w1}. Dispersion is normal, $D_{2a}<0$, see Fig.~\ref{f0} for other parameters.}
	\lab{f5}
\end{figure}

\vspace{1mm}\noindent{\bf Method of slowly varying amplitudes.} 
Here we assume a condition that is quite common for the frequency conversion experiments in $\chi^{(2)}$ microresonators. If a resonator is made to operate close to the $\mu=0$  phase-matching, i.e., $|\ep_0|\sim\kappa_{\zeta}$, then the simultaneous control of the repetition-rate difference between the pump and signal, $D_{1b}-D_{1a}$, is hard to achieve. Therefore, $\mu(D_{1b}-D_{1a})$ easily becomes the dominant frequency scale  in Eq.~\bref{tp1}, i.e.,
$\mu|D_{1a}-D_{1b}|\gg |\delta_{0 b}|,\kappa_\zeta$, $|\ep_0|$, $\gamma_a b_0$. For example, for a bulk-cut ($R\sim 1$mm)~\cite{opo4} and integrated ($R\sim 100\mu$m)~\cite{opo3} LiNbO$_3$ resonators, $(D_{1b}-D_{1a})/\kappa_\zeta\sim 10^3$ and $\sim 10$, respectively. Thus, the above conditions work very well for the former starting from $|\mu|=1$ and for the latter from $|\mu|\sim 10$.

Now, the natural methodological step is to separate the fast and slow time scales in the pump sidebands,
\be
b_{\mu}= B_{\mu}e^{-i\mu(D_{1 b}-D_{1 a})t},~ \mu\ne 0,
\lab{bb}
\ee
where $B_{\mu}$ are the slowly varying amplitudes.
Then,  the $\mu=0$  part of Eq.~\bref{tp1} becomes
\be
\begin{split}
		i\p_t a_{0 }=&\kappa_a\vD_{0a}a_{0} -\gamma_a b_{0} a^*_{0}\\	
		-&\gamma_a\sum_{\mu_1\ne 0}B_{\mu_1}
	a^*_{\mu_1}e^{-i\mu_1(D_{1 b}-D_{1 a})t},
		\\
	i\p_t b_{0}=&\kappa_b\vD_{0b}b_{0} + \frac{i\kappa_b}{2}
	\cH -\gamma_ba_{0}^2	-	\gamma_b\sum_{\mu_1\ne 0}a_{\mu_1}a_{-\mu_1},
	\end{split}
\lab{tp9}
\ee
and the $\mu\ne 0$ part is 
\be
\begin{split}
	i\p_t a_{\mu}=&\kappa_a\vD_{\mu a}a_{\mu} 
	-\gamma_a b_{0}a^*_{-\mu} -B_{\mu}a^*_{0}e^{-i\mu(D_{1 b}-D_{1 a})t}\\
	-&\gamma_a\sum_{\mu_1,\mu_2\ne 0}\wh\delta_{\mu,\mu_1-\mu_2}
	B_{\mu_1}a^*_{\mu_2 }e^{-i\mu_1(D_{1 b}-D_{1 a})t},
 \\
	i\p_t B_{\mu}=&\kappa_b\vD_{\mu b}B_{\mu} -	
	\gamma_b\sum_{\mu_1\mu_2}\wh\delta_{\mu,\mu_1+\mu_2}a_{\mu_1}a_{\mu_2}
	e^{i\mu(D_{1 b}-D_{1 a})t}.
\end{split}
\lab{tp3}
\ee
Here,  $\vD_{\mu\zeta}$ are the auxiliary dimensionless detuning parameters,
\be
\vD_{\mu\zeta}=\left(
\delta_{0 \zeta}+\frac{1}{2}\mu^2D_{2\zeta}-i\frac{1}{2}\kappa_\zeta\right)
\frac{1}{\kappa_\zeta},
\lab{fiel}
\ee
which include the losses and, hence, are complex-valued.
We note that $\vD_{\mu\zeta}$ are free from $D_{1b}-D_{1a}$ 
which has been absorbed by the fast oscillating exponents, see Eq.~\bref{bb}.

Integrating the $\p_t B_\mu$ equation, while assuming that $a_{\mu}$ 
is a slow function of time, we  express the pump sidebands via the signal ones,
\be
B_{\mu}\approx\gamma_b\frac{e^{i\mu(D_{1 b}-D_{1 a})t}}{\mu(D_{1 b}-D_{1 a})}
\sum_{\mu_1\mu_2}\wh\delta_{\mu,\mu_1+\mu_2}
a_{\mu_1}a_{\mu_2}.
\lab{cas}
\ee
Substituting Eq.~\bref{cas} into Eqs.~\bref{tp9} and \bref{tp3} 
would make up the Kerr-like nonlinear terms. 
These terms represent the so-called cascaded Kerr nonlinearity~\cite{op1}, which is, however, negligible in the leading order, 
because it scales inversely with $\mu(D_{1b}-D_{1a})$. Hence, 
Eqs.~\bref{tp9}, \bref{tp3}, and the whole of the master system, Eq.~\bref{tp1}, simplify to
\be
\begin{split}
	&i\p_t a_{0 }=\kappa_a\vD_{0  a}a_{0 }-\gamma_ab_{0 }a^*_{0},
\\	
	&i\p_t a_{\mu }=\kappa_a\vD_{\mu  a}a_{\mu } 	
	-\gamma_ab_{0 }a^*_{-\mu },~\mu\ne 0,
\\
	&i\p_t a_{-\mu }=\kappa_a\vD_{\mu  a}a_{-\mu} 	
	-\gamma_a b_{0 }a^*_{\mu },
\\
	&i\p_t b_{0 }=\kappa_b\vD_{0 b}b_{0 }+ \frac{i\kappa_b}{2}
	\cH-\gamma_ba_{0 }^2 -	2\gamma_b\sum_{\mu_1> 0}a_{\mu_1}a_{-\mu_1}.
\end{split}
\lab{p9}
\ee

Thus,  the pump sidebands,  
$b_{\mu\ne 0}$,  play no significant role in the frequency conversion when the repetition rate difference, $\mu(D_{1b}-D_{1a})$, is large. 
The latter simply is not featured in Eq.~\bref{p9}.
The pumped mode, $b_0$,  is driven by the sum-frequency processes of the 
$\pm\mu$ signal sidebands, which feeds back to 
the equations for the signal sidebands $a_{\pm\mu}$ via the $b_0$-terms. The absence, in the leading order, of the  
sum-frequency interaction between the sidebands with $|\mu_1|\ne|\mu_2|$ 
hints that the generation of the isolated sideband pairs should 
be a preferential regime over the broadband frequency combs.

Our approach is different from, e.g., the method when the whole of the high-frequency field is adiabatically eliminated by one way or the other so that the low-frequency field becomes driven by the cascaded Kerr effect, see, e.g.,~\cite{taki,leo1,op1}. The transition from Eq.~\bref{tp1} to Eq.~\bref{p9} reduces the phase-space dimensionality of the pump field to one, but does not eliminate it entirely, and retains the leading order quadratic nonlinearity.

\vspace{1mm}\noindent{\bf Solutions and  thresholds.}
The reduced model, Eq.~\bref{p9}, allows examining in details 
the properties of the  OPO states (this section) 
and  studying their instabilities with respect to each other (next section).
In what follows we keep using $\mu$ as the running sideband index 
and $\nu$ designates a specific OPO state. 
Fixing $\p_t a_{\pm\nu}=\p_t b_0=0$, and after some engaging 
algebra with Eq.~\bref{p9}, the explicit solutions for the non-degenerate 
OPO states are found in the following form,
\be
\begin{split}
	&a_{\nu}=|a_{\nu}|e^{i\phi_\nu},~
	a_{-\nu}=|a_{\nu}|,~b_{0}=\frac{e^{i\phi_\nu}
	\kappa_a\vD_{\nu a}}{\gamma_a},\\
	&e^{i\phi_\nu}=
	\frac{-i H}
	{\vD_{\nu a}\vD_{0 b}-q_\nu\dfrac{\gamma_a\gamma_b}{\kappa_a\kappa_b}|a_{\nu}|^2},~H=\frac{\gamma_a\cH}{2\kappa_a}, 
\end{split}
\lab{si}
\ee
where, 
$H$ is the dimensionless pump parameter, and $q_\nu=2$ for $\nu\ne 0$. 
Eq.~\bref{si} with $\nu=0$ also covers for the degenerate case, but $q_0=1$. 
Two possible solutions for the sideband amplitudes 
are found by taking modulus squared of the equation for $e^{i\phi_\nu}$,
\ba
\nn
&&|a_{\nu}^\pm|^2	=\frac{\text{Re}(\vD_{\nu a}\vD_{0 b})
	\pm\sqrt{H^2-H_\nu^2+\text{Re}^2(\vD_{\nu a}\vD_{0 b})}}
{q_\nu\gamma_a\gamma_b/\kappa_a\kappa_b},
\\
&& H^2_\nu=\text{Im}^2(\vD_{\nu a}\vD_{0 b})+\text{Re}^2(\vD_{\nu a}\vD_{0 b}).\lab{x3a}
\ea
The recent microresonator theories~\cite{opo3,ol,pra3} reported  the double-valued solutions for the degenerate case. The non-degenerate case was also considered in~\cite{opo3}, but only for the zero detunings and exact phase-matching, while Eq.~\bref{si} and stability analysis that follows 
depend critically on the full account and ability to vary both of those flexibly.

The bifurcation points from the no-OPO to OPO regimes, i.e., Benjamin-Feir instabilities, 
are conditioned by the zeros of the sideband powers,  
\be
\begin{split}
	& |a_\nu^+|^2=0,~\text{i.e.},~H^2= H^2_{\nu},~
	\text{Re}(\vD_{\nu a}\vD_{0 b})<0,~\text{or}\\
& |a_\nu^-|^2=0,~\text{i.e.},~H^2= H^2_{\nu},~
\text{Re}(\vD_{\nu a}\vD_{0 b})>0.
\end{split}
\lab{q0}
\ee
The laser power $\cW_\nu$ at the OPO threshold is then calculated from
$\cW_\nu=8 \pi\kappa_a^2 H_\nu^2/\cF\gamma_a^2$.
The left column in Fig.~\ref{f2} shows  $\cW=\cW_\nu$   vs $\delta_{0 b}$,  for 
the negative, zero, and positive phase-mismatch $\ep_0$. 
The minimum of $\cW$ near $\delta_{0 b}=0$ exists in all three cases. 
For $\ep_0\le 0$, this minimum is provided by the $\nu=0$ mode, i.e.,
the no-OPO state losses its stability to the degenerate OPO first.
The second minimum of $\cW$ at $\delta_{0 b}= -\ep_0$, 
i.e., $\delta_{0a}=0$ (see Eq.~\bref{fi1b}), also happens for $\nu=0$. 

The top row of Fig.~\ref{f3} shows the results of numerical simulations
of Eq.~\bref{tp1} across the regions of the unstable no-OPO state 
for the relatively small input powers marked with the black dashed lines in the first column of Fig.~\ref{f2}. The  $\ep_0=0$ case in Fig.~\ref{f3}a  demonstrates how the degenerate OPO is first excited from the no-OPO state and then switches, in a cascaded manner, to the non-degenerate OPOs. 
Fig.~\ref{f3}b shows how this scenario can happen twice for $\ep_0<0$. 
We note that the two cascades in Fig.~\ref{f3}b occur in the reverse order.
The right-to-left cascade involves oscillations of the $\mu=0$ and other modes (breathing comb), while all the left-to-right cascades in Figs.~\ref{f3}a-\ref{f3}c appear as the direct transitions between the neighbouring non-degenerate OPO states, as it is expected to happen in the Eckhaus instability scenario.
We recall that we consider the near-phase-matching between the $\mu=0$ modes in the pump and signal fields. It leads to the  Rabi-like oscillations between the $a_0$ and $b_0$ modes, which acquire small gain in the narrow interval of detunings and give birth to the breather states, see Refs.~\cite{prr,pra2} for details of the theory of the $\chi^{(2)}$ Rabi oscillations.

For $\ep_0>0$, the minimum of $\cW$ is also  found at $\delta_{0b}\approx 0$, 
but now it happens for $\nu\ne 0$,   see Fig.~\ref{f2}b.
Thus, here, the no-OPO state transits directly to the non-degenerate regime.  Fixing $\delta_{0b}=0$ in Eq.~\bref{x3a} gives $16H_{\nu}^2=\ep_\nu^2/\kappa_a^2+1$, where $\ep_\nu=\ep_0+\nu^2 D_{2a}$.
Hence, the minimum threshold power, $H_{\wh\nu}^2=1/16$, for the Benjamin-Feir instabilities is achieved at the exact phase-matching, $\ep_{\wh\nu}=0$, where $\wh\nu\approx\sqrt{-\ep_0/D_{2a}}$, $\ep_0 D_{2a}<0$.
Fig.~\ref{f3}c~demonstrates how the below threshold OPO~switches directly 
to the non-degenerate state and passes through the cascade of~$\nu$'s.

Solving $H^2=H^2_\nu$, Eqs.~\bref{q0}, one could find either two or four real values of $\delta_{0b}$ where the $\pm\nu$ sidebands bifurcate from zero, cf.,  Figs.~\ref{f3}d, \ref{f3}f
with \ref{f3}e. 
In the left column of Fig.~\ref{f2}, the different colours mark the parts of the $H^2=H_\nu^2$ thresholds where either $|a_\nu^+|^2=0$ (red) or $|a_\nu^-|^2=0$ (blue). The points of transition between the two colors are found by setting 
\be
\text{Re}(\vD_{\nu a}\vD_{0 b})=0.
\lab{w1}
\ee
Tuning the pump laser across the red boundaries and moving into the instability interval leads to   $|a_\nu^+|^2$ gradually increasing from zero, which corresponds to the soft-excitation regime (supercritical bifurcation), see Fig.~\ref{f3}d. 
Entering the instability tongue across the blue boundary leads to $|a_\nu^+|^2$ popping out stepwise and
$|a_\nu^-|^2$ bifurcating from zero subcritically (hard excitation), see Figs.~\ref{f3}e, \ref{f3}f. 

The right column in Fig.~\ref{f2} shows the existence and stability ranges
of the degenerate, $\nu=0$, OPO states. They bifurcate from the no-OPO state super-critically along the red line and sub-critically from the blue line. 
The parameter range where the $|a_0^+|^2$ and $|a_0^-|^2$ solutions 
coexist is located between the blue and dashed-grey lines. 
Generalising for arbitrary $\nu$, the dashed-grey boundary is found from
\be
|a_\nu^+|^2=|a_\nu^-|^2, \text{~i.e.,~}
H^2=\text{Im}^2(\vD_{\nu a}\vD_{0b}).
\lab{lim}
\ee

\vspace{1mm}\noindent{\bf Eckhaus instabilities and OPO tuning.}
While the top row in Fig.~\ref{f3} shows the sequential switching between the 
modes with different numbers and intervals of the pump detuning that select 
a particular sideband pair, the bottom row shows the changes of the sideband amplitudes computed from Eq.~\bref{tp1} and maps them on the  analytical solutions for the OPO states.  
Apart from the oscillatory instabilities around $\delta_{0b}/\kappa_b\approx 0.5$ in Figs.~\ref{f3}b, \ref{f3}e, all the insatiabilities of  a given OPO state converge to the nearby stable one. In other words, these instabilities lead to the $\nu\to\nu+1$ swaps and, hence, to the change of the spatial period and frequency of the waveform in the resonator, i.e., these are the discrete spectrum Eckhaus instabilities.

By taking the OPO state with an arbitrary $\nu\geqslant 0$, adding small perturbations
$\hat a_\mu(t)$, where $|\mu|\ne\nu$, and linearising the  
slowly-varying-amplitude model~\bref{p9} we find 
\be
\begin{split}
	&i\p_t\hat  a_{\mu }=\kappa_a\vD_{\mu  a}\hat a_{\mu } 	
	-\gamma_ab_{0 }\hat a^*_{-\mu },
	\\
	&i\p_t\hat  a_{-\mu }=\kappa_a\vD_{\mu  a}\hat a_{-\mu} 	
	-\gamma_a b_{0 }\hat a^*_{\mu },
\end{split}
	\lab{l9}
\ee
where $b_0$ is a function of $\nu$ in accordance with Eq.~\bref{si}.
Solving Eq.~\bref{l9} with $\hat a_{\mu }(t)=\tilde a_{\mu }\exp\{t\lambda_{\nu,\mu} \}$ and 
$\hat a_{-\mu }^*(t)=\tilde a_{-\mu }\exp\{t\lambda_{\nu,\mu} \}$ yields a set of  the sideband-pair growth rates, 
\be
\begin{split}
\lambda_{\nu,\mu}=
&-\tfrac{1}{2}\kappa_a+\kappa_a\sqrt{
	|\vD_{\nu a}|^2-|\vD_{\mu a}|^2}\\
=&-\tfrac{1}{2}\kappa_a+\sqrt{
	\delta_{\nu a}^2-\delta_{\mu a}^2}.
\end{split}
\lab{th1}
\ee
Thus, Eqs.~\bref{th1} describe the growth rates of the Eckhaus instabilities in the microresonator OPO,
i.e., destabilization of the non-degenerate OPO state corresponding to the $\pm\nu$ sideband pair through the excitation of any other pair $\pm\mu$.
The instability threshold is reached when $\lambda_{\nu,\mu}$
becomes zero, while the generation of the $\pm\nu$ sideband pair is stable if the pump frequency is tuned to provide $\delta_{0b}$ such that $\delta_{\nu a}^2\leqslant\delta_{\mu a}^2+\tfrac{1}{4}\kappa_a^2$. The oscillatory bifurcations, i.e., the birth of breathers highlighted by the white lines in Fig.~\ref{f2}, correspond to $\nu=\mu$, and, therefore, are exempt from the above theory.

To find conditions of the switching from one sideband pair to the next, 
we set $\mu=\nu\pm 1$. For $D_{2a}<0$ (normal dispersion), 
the interval of stable generation of the $\pm\nu$ sideband pair is found as
\ba
-\tfrac{1}{2}D_{2a}[\nu^2+&&(\nu- 1)^2]-\ep_-
\leqslant
\nn\\
&& \delta_{0b}\leqslant 
-\tfrac{1}{2}D_{2a}[\nu^2+(\nu+ 1)^2]-\ep_+,
\lab{in1}
\ea 
where $\ep_\pm=\ep_0+\kappa_a^2/2D_{2a}(1\pm 2\nu)$. The instability boundaries, $\lambda_{\nu,\nu\pm 1}=0$, as given by the above condition, 
are shown in~Figs.~\ref{f3},~\ref{f4}a by the white dashed vertical lines
and they match perfectly with the transitions found from the
numerical modelling of Eq.~\bref{tp1}. 

The detuning corresponding to the midpoint 
of every step on the Eckhaus ladder, i.e., the left plus right 
limits divided by two, is  $\delta_{0b}\approx
-\ep_0-D_{2a}(\nu^2+\tfrac{1}{2})$. 
Thus, the latter reproduces the characteristic parabolic shape of  $\delta_{0b}$ vs $\nu$ 
in Fig.~\ref{f4}a. Recalling Eqs.~\bref{om}, \bref{ep}, 
one can show that the same parabola is described by 
\ba
\delta_{0b}=-\ep_\nu-\tfrac{1}{2} D_{2a},
\lab{in2}
\ea
Thus, the analysis done so far has demonstrated that a sequence of the OPO regimes achieved by the scan of the pump frequency follows the steps of the Eckhaus instabilities ladder and the phase-matching condition, cf., Eqs.~\bref{ep} and \bref{in2}. It also reveals that the size of the Eckhaus ladder's steps and the tuning curve's discreetness are controlled by the second-order dispersion. We shell note that the applicability of Eq.~\bref{th1} extends  beyond the second-order expansion applied for $\om_\mu$, Eq.~\bref{om}, while Eq.~\bref{in2} relies on it. Therefore, the microresonators with dispersion dominated by the higher orders could be an interesting case to consider in future.

Fig.~\ref{f4}a shows a sequence of the OPO transitions for 
the power which is much higher than in Fig.~\ref{f3} (cf. black and white lines in Fig.~\ref{f2}b). 
The  Eckhaus ladder in Fig.~\ref{f4}a is shown up to $\nu=30$, but extends upto $\nu=\nu_\text{max}= 120$. 
The data in the top row of Fig.~\ref{f3} also demonstrate that, for $\ep_0>0$, 
the parabolic tuning curve is also cut at its minimum.
The power reduction, from the levels in Fig.~\ref{f4}a to the ones in Fig.~\ref{f3}, 
does not change the shape of parabola but reduces its extent in $\nu$ and $\delta_{0b}$.
Since the value of $\nu_\text{max}$ sets the practical limit for the OPO tunability, it 
is now mandatory to apply our theory to elaborate this point further.

First, we should recall that the data in Fig.~\ref{f3} have confirmed that 
Eqs.~\bref{si}, \bref{x3a} provide an excellent approximation for the OPO states.
Fig.~\ref{f4}d includes the much weaker, practically negligible, part of the spectrum and, thereby, explicitly reveals the degree of accuracy of Eqs.~\bref{si}, \bref{x3a}. It also uncovers that the non-degenerate OPO states belong to the family of staggered combs.
It follows from Eq.~\bref{si} that the power of the signal sidebands, $a_{\pm\nu}$, depends on the laser power,  $H^2\sim\cW$, and the power of $b_0$ does not, and only its phase does. 
Indeed, $b_{0}^\pm=e^{i\phi_\nu^\pm}\kappa_a\vD_{\nu a}/\gamma_a$, where $\phi_\nu^\pm=\phi_\nu(|a_\nu^\pm|^2)$.
This is a reason why $H^2$ does not explicitly enter the Eckhaus instability rate, see  Eq.~\bref{th1}.  However, the limits of  existence of $|a_{\nu}^\pm|^2\geqslant 0$, and, hence, of $b_{0}^\pm$ in   Eq.~\bref{l9} do critically depend on $H^2$ as is given by Eq.~\bref{lim}. 

Substituting the detunings $\delta_{0b}=-\ep_\nu$ (corresponding to the numbered tips in the left column of Fig.~\ref{f2}) in 
Eq.~\bref{lim} yields $4\kappa_b^2H^2=(\ep_0+\nu^2 D_{2a})^2$. From here one can work out the power dependencies of the minimal, $\nu_{\min}$ vs $\cW$, and maximal, $\nu_{\max}$ vs $\cW$,
mode numbers corresponding to the $\pm\nu$ OPO states. 
Recalling that the dispersion is normal, $D_{2a}<0$, the maximal number is 
\be
\nu_\text{max}^2\approx\frac{-\ep_0- 2\kappa_bH}{D_{2a}}=
\frac{-\ep_0}{D_{2a}}
-\frac{\kappa_b\gamma_a}{\kappa_aD_{2a}}\sqrt{\frac{\cW\cF}{2\pi}}.
\lab{ww}
\ee
The $\nu_{\min}$ is conditioned by the sign of $\ep_0$. It is zero for $\ep_0\leqslant 0$
and $\nu_\text{min}^2\approx (-\ep_0+ 2\kappa_bH)/D_{2a}$ if $\ep_0>0$, see Fig.~\ref{f3}. 
The plots of $\nu_{\max}$, $\nu_{\min}$ vs the laser power are shown in Fig.~\ref{f5}. 

For the whispering gallery and integrated LiNbO$_3$ resonators, the mode number change by one corresponds to the wavelength step $\sim 0.1$nm~\cite{opo4} and $2$nm~\cite{opo2}, respectively. Hence $\nu_\text{max}=50$ corresponds to the wavelength difference between the signal and idler 10nm and 200nm, respectively. The pump frequency tuning data in Ref.~\cite{opo2} report up to $200$nm signal-idler separations. The experimental reports of the parabolic pump-frequency tuning curves in the microresonator~\cite{opo1,opo3,opo2}, fiber-loop~\cite{mar}, 
and bow-tie~\cite{leo1} OPOs do not contain the  data sufficient for making a comparison with the reported here power dependent parabola cut-off at $\nu_{\max}$ and $\nu_{\min}$. Therefore, further research into this problem is necessary.

\vspace{1mm}\noindent{\bf Conclusions}\\
We have demonstrated that the degenerate OPO transiting to the ladder of the nondegenerate OPO states via a sequence of Eckhaus instabilities is a universal scenario if the repetition rate difference between the signal and pump fields, $|D_{1b}-D_{1a}|$, dominates over all other frequency scales. The emerging combs are most typically staggered,  with the two dominant sidebands in the signal field, which corresponds to the nondegenerate OPO regime, see Figs.~\ref{f0},~\ref{f3}  and \ref{f4}. For the high-finesse whispering gallery resonator used as an example in our work, this scenario covers the span of the pump laser powers from nano to milli Watts. The more complex frequency combs start to appear for the pump powers above $3$mW, see $\delta_{0b}\approx -20\kappa_b$ in Figs.~\ref{f4}a and \ref{f4}b.  
Apart from increasing the pump power, the ways to facilitate the microresonator OPOs to operate in the non-staggered comb could be, e.g., using resonators with the better matched repetition rates taken relative to the losses, i.e., having smaller $|D_{1b}-D_{1a}|/\kappa_{a,b}$, and using pumping into the modes with the odd numbers~\cite{tang1,op1,pra3,ol}.

The slowly-varying amplitude model, Eq.~\bref{p9}, that we derived and employed here was the key step that has allowed us to solve the problem of switching between the OPO operating in the $\pm\nu$
sideband pair to any other mode pair $\pm\mu$, where $\nu\ne\mu$. 
This model relies on the aforementioned dominance of the repetition rate difference, $|D_{1b}-D_{1a}|$,
which drives the fast oscillations of the pump sidebands, see Eq.~\bref{bb}. It also reveals that the growth rate of the Eckhaus instability does not depend on $|D_{1b}-D_{1a}|$ in the leading order, see Eq.~\bref{th1}.
Regarding the quadratic nonlinearity, the slowly-varying amplitude model highlights the dominant role of the terms responsible for the nonlinear mixing of the central mode of the pump field with all the signal sidebands. 

Our methodology has also let us derive the equation for the maximally allowed sideband order generated by the pump frequency tuning method in the microresonator OPOs, which is easy to apply and use as a practical guideline for the parametric down-conversion experiments. All our analytical results are in excellent agreement with the numerical modelling of the full coupled-mode system, see Eq.~\bref{tp1}.

\vspace{1mm}\noindent{\bf Acknowledgements}\\
We acknowledge the financial support received from  UK EPSRC (2119373, DTP studentship) and EU Horizon 2020 Framework Programme (812818, MICROCOMB).

\vspace{1mm}\noindent{\bf Author contributions}\\
D.N. Puzyrev developed numerical codes, performed simulations, analysed data and prepared them for publication. D.V. Skryabin developed the theory and wrote the text.

\vspace{1mm}\noindent{\bf Competing interests}\\
The authors declare no competing interests.
D.V.~Skryabin is a Guest Editor of the Collection “Microresonator Frequency Combs: New Horizons” for Communications Physics, but was not involved in the editorial review of, or the decision to publish this article.

\vspace{1mm}\noindent{\bf Data availability}\\
The data supporting the findings of this study are available from authors on reasonable request.

\vspace{1mm}\noindent{\bf Code availability}\\
The codes used for this study are available from the first author 
(dp710@bath.ac.uk) upon reasonable request.

\vspace{2mm}\noindent{\bf METHODS}\vspace{2mm}\newline
\noindent 
Numerical integration of the coupled-mode model, see Eq.~\bref{tp1}, has been performed using a fourth-order Runge–Kutta method. The procedure was optimized applying the fast Fourier transform algorithm to the nonlinear terms in the real space to compute 
the sums in Eq.~\bref{tp1}, see Ref.~\cite{josab} for details. The data shown in Figs.~\ref{f3}, and \ref{f4}  were obtained for 256 optical modes in the 'a' and 'b' fields, i.e., for $\mu=-127,\dots 128$. The initial conditions were the no-OPO state, see Eq.~\bref{p1}, plus random noise in all modes.

\end{document}